\documentclass[12pt]{article}
\usepackage{amssymb}
\usepackage{enumerate}
\usepackage{dcolumn}

\usepackage{graphicx}
\usepackage{dcolumn}
\usepackage{bm}

\begin{document}


\def\H{{\cal H}}
\def\ttheta{\tilde{\theta}}

\def\beq{\begin{equation}}
\def\eeq{\end{equation}}
\def\bea{\begin{eqnarray}}
\def\eea{\end{eqnarray}}
\def\ben{\begin{enumerate}}
\def\een{\end{enumerate}}
\def\la{\langle}
\def\ra{\rangle}
\def\a{\alpha}
\def\b{\beta}
\def\g{\gamma}\def\G{\Gamma}
\def\d{\delta}
\def\e{\epsilon}
\def\phi{\varphi}
\def\k{\kappa}
\def\l{\lambda}
\def\m{\mu}
\def\n{\nu}
\def\o{\omega}
\def\p{\pi}
\def\r{\rho}
\def\s{\sigma}
\def\t{\tau}
\def\S{\Sigma }
\def\gsim{\; \raisebox{-.8ex}{$\stackrel{\textstyle >}{\sim}$}\;}
\def\lsim{\; \raisebox{-.8ex}{$\stackrel{\textstyle <}{\sim}$}\;}
\def\gtrsim{\gsim}
\def\lessim{\lsim}
\def\loc{{\rm local}}
\def\vm{v_{\rm max}}
\def\bh{\bar{h}}
\def\del{\partial}
\def\nab{\nabla}
\def\half{{\textstyle{\frac{1}{2}}}}
\def\fourth{{\textstyle{\frac{1}{4}}}}

\def\L{\Lambda}
\def\l{\lambda}
\def\g{\gamma}
\def\SL{S_{\rm{LQG}}}
\def\SB{S_{\rm{BH}}}
\def\Geff{G_{\rm eff}}
\def\Aeff{A_{\rm eff}}
\def\Leff{\L_{\rm eff}}
\def\Gm{G_{\rm micro}}
\def\Am{A_{\rm micro}}
\def\Lm{\L_{\rm micro}}

\begin{center} {\Large \bf
 Renormalization and black hole entropy\\ in Loop Quantum Gravity}
\end{center}

\vskip 5mm
\begin{center} \large
{{Ted Jacobson\footnote{E-mail: jacobson@umd.edu}}}\end{center}

\vskip  0.5 cm
{\centerline{\it Department of Physics}}
{\centerline{\it University of Maryland}}
{\centerline{\it College Park, MD 20742-4111, USA}}

\vskip 1cm

\begin{abstract}
Microscopic state counting for a black hole in Loop
Quantum Gravity yields a result proportional to
horizon area, and inversely proportional to Newton's
constant and the Immirzi parameter. It is argued here
that before this result can be compared to the
Bekenstein-Hawking entropy of a macroscopic black
hole, the scale dependence of both Newton's constant
and the area must be accounted for. The two entropies
could then agree for any value of the Immirzi
parameter, if a certain renormalization property
holds.

\end{abstract}


The number of microscopic states of a black hole has been computed
in Loop Quantum Gravity (LQG), in the state space of spin networks.
The result for the entropy of a black hole with horizon area $A$ is
\beq S_{\rm {LQG}}= \frac{b}{\gamma}\frac{A}{\hbar G}, \label{SL}\eeq
where $b$ is a numerical constant and $\g$ is the
Immirzi parameter. These calculations have a long and
continuing history (see for example \cite
{Krasnov:1996tb,Rovelli:1996dv,Ashtekar:1997yu,
Domagala:2004jt, Meissner:2004ju, Dreyer:2004jy,
Ghosh:2004wq, Tamaki:2005jp,Corichi:2006bs,Tamaki:2007iy}
and for reviews
\cite{Ashtekar:2004eh,
Thiemann:2006cf,Corichi:2007zz}), including some
controversy over the correct evaluation of the number
of states. The results differ only in the value of $b$
however (unless states related by surface
diffeomorphisms are identified, as has discussed for
example in \cite{Alekseev:2000hf}). In addition to the
case of spherically symmetric, static black holes, the
result (\ref{SL}) has been shown to hold, with the
{\it same} value of $b$, in the presence of scalar,
Maxwell, and Yang-Mills fields~\cite{Ashtekar:2000eq}
as well as for spinning black holes in pure
GR~\cite{Ashtekar:2004nd}. According to current
understanding, the Immirzi parameter represents a
quantization ambiguity in LQG, which affects physical
observables. That is, it is generally believed that
different values of $\g$ define physically different
quantum theories.

Most discussions of the entropy  observe that
in order for $\SL$ to match the Bekenstein-Hawking entropy $S_{\rm
BH}=A/4\hbar G$, the Immirzi parameter must be chosen to have a
special value,
\beq \g=4b. \label{g1}\eeq
Not much is said about {\it why} $\g$ should have this particular
value, but the idea seems to be that one can regard this calculation
as a {\it derivation} of the ``correct" value. Since any
Immirzi parameter may be adopted in the microscopic theory,
such an interpretation can make sense only if, as suggested
for example in  \cite{Ashtekar:2004eh}, this is the unique
value for which the theory possess the classical limit of
general relativity.

While it may be a logical possibility that the
existence of a classical GR limit requires a unique
value of $\gamma$, I can see no reason other than the
entropy calculation to suspect that. Moreover, it is
hard to see how the entropy computations carried out
so far could probe LQG deeply enough to ascertain such
a consistency condition for the continuum limit. The
reason is that these computations involve only the
kinematics, not the {\it dynamics} of the theory.
Thus, it seems to me more reasonable to expect that,
if the microscopic calculation of black hole entropy
is really correct, then it should agree with $S_{\rm
BH}$ for {\it all} values of $\g$.

On the face of it this looks impossible, however in
fact the comparison of $\SL$ with $A/4\hbar G$ has
been made prematurely. The latter refers to a property
of a semiclassical black hole. As such, the area is
measured using the low energy effective metric field,
and the Newton constant is the low energy effective
Newton constant. By contrast, the area and Newton
constant in $\SL$ are the microscopic quantities
appearing in the fundamental formulation of the theory
\cite{history}. This raises the question of exactly
how the ``correspondence principle" between the
microscopic description in LQG and the effective field
theory (EFT) description operates.

Anything said about the correspondence now is
necessarily provisional, since fundamental aspects of
LQG have not yet been understood. The Hamiltonian
constraint, which encodes all the dynamics, remains to
be understood, much less solved. Once solved, the
theory can in principle only make statements about
diffeomorphism invariant observables, and only
tentative first steps have been achieved for the
identification of a suitable class of such
observables. Nevertheless, the setting of the black
hole entropy computations is well-defined, and
reasonably well-motivated, at least enough to merit
scrutiny of its theoretical basis and
self-consistency.

We now come to the central point of this note: there
is no reason to expect the correspondence between
microscopic and macroscopic quantities to be trivial.
There are two different reasons for this. One is that
in translating from a discrete spin-network to a field
theoretic framework there is a profound change of
objects and language.  The other, independent reason
is that even once the correspondence to an effective
field theory has been made, the renormalization group
flow of {\it that} theory from the UV to the IR is
highly nontrivial. What then can be said?

Let us initially consider pure gravity, for
simplicity, and let us assume that the microscopic
cosmological constant $\Lm$ is zero and that an EFT
limit exists. Dimensional analysis alone then implies
that the low energy effective couplings are related to
the microscopic parameters via
\beq \Leff=f(\g)/\hbar \Gm,
\nonumber\qquad\Geff=g(\g)\Gm,\label{rencouplings}
\eeq
where $f(\g)$ and $g(\g)$ are at this stage unknown
functions of the Immirzi parameter.

The relation between the areas assigned to a given
surface, at the macro and micro levels of description,
is only constrained by dimensional analysis to have
the form
\beq
 \Aeff=k(\g,\Am/\hbar \Gm)\Am,
 \label{ren}  \eeq
where $k$ is an unknown function of its dimensionless
arguments. However, since the area should be additive
with respect to a decomposition of the surface into
parts, it seems that $\Aeff$  must scale linearly with
$\Am$. Moreover, if the areas are {\it not}
proportional, then the microscopic entropy does not
scale with the macroscopic area, so cannot possibly
agree with $\SB$. Since I merely want to ask if it is
theoretically {\it possible} that the entropies do
agree for all values of $\g$, I will therefore assume
the areas are related as
\beq
 \Aeff=h(\g)\Am.
   \label{renarea}\eeq
The notion of area renormalization may at first appear
alien, but note that the macro and micro ``structure"
of a surface are in principle quite different notions.
(For an approach to the notions of coarse-graining and
area renormalization in LQG see \cite{Livine:2005mw}.)
Moreover, on the EFT level, one expects a nontrivial
relation between the low energy effective metric field
and the metric at a UV cutoff scale (see e.g.\
\cite{Reuter:2006zq} and references therein).

Although $f$, $g$ and $h$ relate quantities in the
discrete and continuous frameworks, they are similar
to the usual renormalization constants of QFT, so I
will call them by that same name. The relation for the
effective cosmological constant $\Leff$ has been
indicated in (\ref{rencouplings}), but I will say no
more about it here except to assume that $\hbar
G_{\rm{eff}} \L_{\rm{eff}} \ll 1$, i.e. that the
cosmological constant is very small in units of the
effective Planck length, so that it may be ignored
from here on.

In writing these relations, and in the rest of this
paper, I am simply {\it assuming}
that a semiclassical EFT limit of LQG exists, for at least some
$\g$.
The problem of understanding these renormalization
relations is a crucial part of understanding the
semiclassical limit of LQG. While this is a wide open
problem, it can nevertheless be asked at present
what properties must they exhibit if the
existing microscopic black hole entropy calculations
are to be valid?

The Bekenstein-Hawking entropy $\SB$ of a macroscopic
black hole is expressed
in terms of the ``renormalized" quantities as
\beq \SB=\frac{A_{\rm eff}}{4\hbar G_{\rm eff}}, \eeq
while the microscopic LQG result (\ref{SL}) should be written
as
\beq S_{\rm {LQG}}= \frac{b}{\gamma}\frac{\Am}{\hbar \Gm}.
\label{SLm}\eeq
These will agree provided
\beq \gamma = 4b\,
\frac{\Am/\Gm}{A_{\rm{eff}}/G_{\rm{eff}}}. \label{g2}
\eeq
The relation between the effective and microscopic quantities
must therefore be taken into account.

In terms of the renormalization constants defined in
(\ref{rencouplings}) and (\ref{renarea}), condition
(\ref{g2}) becomes
\beq \gamma = 4b\, \frac{g(\g)}{h(\g)}.
\label{g3}
\eeq
Condition (\ref{g3}) may have (i) no solutions, (ii) a
discrete set of solutions, or (iii) a continuous range
of $\g$ might be solutions. Without knowing more about
$g(\g)/h(\g)$, we can not say which of these is the
case. If there are no solutions, then $\SB$ is never
recovered, from which we infer that $\SL$ can not in
fact be the correct black hole entropy in LQG. If
there is instead a discrete set of solutions, we are
back in the position of having to maintain that, for
some mysterious reason, the counting works only for
special values of $\g$. The final possibility is that
a continuous range of values of $\g$ are solutions to
(\ref{g3}).

If a range of values of $\g$ solve (\ref{g3}),
then there must be a tight relation between the
renormalization of area and Newton's constant.
That is, the ratio $g(\g)/h(\g)$ must be proportional to
$\g$, and with the particular coefficient $1/4b$.
Is this conceivable?

If the area is not renormalized,  i.e. if
$h(\g)\equiv1$, then (\ref{g3}) would imply
$g(\g)=\g/4b$. It is not easy to imagine that the
renormalization of Newton's constant is simply
proportional to $\g$, nor that it always has this
form, with the same coefficient, independent of the
matter (Maxwell field, etc.) that is coupled to the
metric. In fact, from the EFT side it is clear that
coupling to different sorts of matter {\it must} alter
the renormalization of $G$. Therefore it appears
untenable that, without area renormalization, the LQG
entropy computation yields $\SB$ for all values of
$\gamma$ (and for generic matter).

Area renormalization significantly changes the picture
however. The renormalization of $G$ can then be
arbitrarily complicated, and dependent on the matter
content, as long as it is universally tied to the
renormalization of the area operator in the
appropriate fashion indicated by (\ref{g3}). I see no
reason why this could not be the case. If the
continuum limit indeed exists, and the LQG black hole
state counting is correct, then I would assert that it
{\it must} be the case.


I have indicated a scenario in which $\SL$ correctly counts
the black hole entropy, without the need to select a special
Immirzi parameter.  It involves an unproven
hypothesis about the correspondence
between LQG and the semiclassical EFT of GR.
Can this hypothesis be tested?

One thing that can be done is to ask whether it could
possibly be robust allowing for additional theory
parameters. For instance, if a microscopic
cosmological constant $\Lm$ is included, then a
dimensionless quantity $\l=\hbar \Gm\Lm$ exists, and
the renormalization constants can depend upon $\l$.
Another example of a parameter would be a gauge
coupling constant. Allowing for some collection of
such parameters would change nothing essential in the
above discussion, so the required robustness seems at
least possible. However, this does make clear that if
the hypothesis is valid, it must be due to a very
general feature of the emergence of the EFT
description.

To directly test the hypothesis requires an improved
understanding of the EFT limit, but need not
necessarily involve black holes. Although the
numerical constant $b$ is the one that arises in the
horizon state counting, it does so in a way that might
be more generally relevant to the relation between
semiclassical geometry of areas and the microscopic,
spin network variables. Thus perhaps the validity of
the relation (\ref{g3}) can be tested in a simpler
setting, for example even in the vacuum.

\section*{Acknowledgments} I am grateful to
Kirill Krasnov, Don Marolf, Lee Smolin, and Aron Wall
for useful comments and discussions on a draft of this
paper. This work was supported in part by the National
Science Foundation under grant PHY-0601800.

\end{document}